\documentclass{article}
\usepackage{epsfig}
\usepackage{amsmath}
\usepackage{amsfonts}\tolerance=10000
\date{\today} 
 
\usepackage{graphicx}

\newcommand{\insertplot}[5]{\begin{figure}
 \hfill\hbox to 0.05in{\vbox to #5in{\vfill 
 \inputplot{#1}{#4}{#5}}\hfill}
 \hfill\vspace{-.1in}
 \caption{#2}\label{#3}
 \end{figure}}
 \newcommand{\inputplot}[3]{
 \special{ps: plotfile #1}
}
\newcounter{fig}

\newcommand{\beq}{\begin{equation}}
\newcommand{\eeq}{\end{equation}}
\newcommand{\beqs}{\begin{eqnarray}}
\newcommand{\eeqs}{\end{eqnarray}}

\numberwithin{equation}{section}
\newcommand{\be}{\begin{equation}}
\newcommand{\ee}{\end{equation}}
\newcommand{\bea}{\begin{eqnarray}}
\newcommand{\eea}{\end{eqnarray}}

\usepackage{graphicx}

\begin{document}

\title{Einstein-Gauss-Bonnet black holes in de Sitter spacetime \\and the quasilocal formalism } 
 
\author{
{\large Yves Brihaye}$^{\dagger}$
and {\large Eugen Radu}$^{\ddagger }$ \\ \\
$^{\dagger}${\small Physique-Math\'ematique, Universite de
Mons-Hainaut, Mons, Belgium}\\
$^{\ddagger}${\small Institut f\"ur Physik, Universit\"at Oldenburg, Postfach 2503
D-26111 Oldenburg, Germany}
  }

\maketitle 
 
\begin{abstract}
We propose to compute  the action and global charges of the asymptotically  de Sitter solutions
in Einstein-Gauss-Bonnet theory  by using the counterterm method in conjunction 
with the quasilocal formalism.
The general expression of the counterterms and the boundary stress tensor is presented for spacetimes
of dimension $d\leq 7$.
We apply this tehnique for several different solutions in Einstein-Gauss-Bonnet theory
with a positive cosmological constant.
Apart from known solutions,  we consider also $d=5$ vacuum rotating black holes
with equal magnitude angular momenta.
These solutions are constructed numerically within a nonperturbative approach, 
by directly solving the Einstein-Gauss-Bonnet equations
with suitable boundary conditions.
\end{abstract}

%%%%%%%%%%%%%%%%%%%%%%%%%%%%%%%%%%%%%%%%%%%%%%%%%%%%%%%%%%%%%%%%%%
\section{ Introduction}
%%%%%%%%%%%%%%%%%%%%%%%%%%%%%%%%%%%%%%%%%%%%%%%%%%%%%%%%%%%%%%%%%%
 One of the most
fruitful approaches in computing conserved quantities 
in general relativity
is to employ the
quasilocal formalism \cite{Brown:1992br}. The basic idea here is to enclose a given
region of spacetime with some surface, and to compute all relevant
(conserved and/or thermodynamic) quantities with respect to that surface. 
For a spacetime that is either asymptotically anti--de Sitter (AdS) 
\cite{Balasubramanian:1999re}, \cite{Emparan:1999pm}, \cite{Henningson:1998gx} or
asymptotically  flat \cite{Kraus:1999di}, \cite{Mann:2005yr}, \cite{Astefanesei:2005ad} 
it is possible to extend the quasilocal
surface to spatial infinity without difficulty, provided one incorporates
appropriate boundary terms in 
the action to remove divergences.
The  boundary terms are built up with
curvature invariants of  the boundary metric 
and thus obviously they do not alter the bulk equations of motion.
Therefore,  this approach has the nice feature that it is not necessary to embed  the  boundary geometry 
in a reference background (see also \cite{Cai:1999xg}-\cite{Kleihaus:2006ee} for other applications of this
formalism for a different asymptotic structure of spacetime).

The situation is much more involved for asymptotically de Sitter (dS) spacetimes,
 because of the absence of spatial infinity and a globally timelike Killing
 vector in this case.  
In the prescription proposed in \cite{Balasubramanian:2001nb}, these obstacles are
avoided by computing the quasilocal
tensor of Brown and York \cite{Brown:1992br} (augmented by suitable boundary counterterms), 
on the Euclidean surfaces at future/past timelike infinity $\mathcal{I}^{\pm}$.  
This allows also a discussion of the thermodynamics
of the asymptotically dS solutions outside the cosmological horizon, 
the boundary counterterms regularising the (tree-level)  gravitation action  as well.
The efficiency of this approach has been demonstrated in a broad range of examples, including configurations
 with gravitating matter fields \cite{Ghezelbash:2001vs}-\cite{Dehghani:2002nt}.
 
The results in \cite{Balasubramanian:2001nb}-\cite{Dehghani:2002nt} concern  the case of Einstein gravity with a positive
cosmological constant. 
However, for a spacetime dimension $d>4$, the Einstein gravity presents a natural generalisation
-- the so called Lovelock theory, constructed 
from vielbein, the spin connection and their exterior derivatives without using the Hodge dual,
such that the field equations are second order  \cite{Lovelock:1971yv}, \cite{Mardones:1990qc}.
Following the Ricci scalar, the next order term in the Lovelock hierarcy is the Gauss-Bonnet (GB) one,
which contains quadratic powers of the curvature.
As discussed in the literature, this term appears as the first curvature stringy
correction to general relativity~\cite{st,Myers:1987yn}, when assuming
that the tension of a string is
large as compared to the energy scale of other variables.  

In principle, there are no obstacles in computing the action and global charges of EGB solutions 
in dS spacetime
by using a quasilocal formalism similar to that proposed in \cite{Balasubramanian:2001nb} for the Einstein gravity.
At any given dimension one can write down only a finite 
number of counterterms that do not vanish at future/past timelike 
infinity.
This feature does not depend upon the bulk theory is Einstein or GB.
However, the presence in this case of a new length scale (the GB coupling constant) implies a complicated
expression for the coefficients of the boundary  counterterms and makes the procedure
technically much more involved.

The corresponding problem for an asymptotically AdS spacetime has been discussed 
in the recent paper \cite{Brihaye:2008xu}
(see also \cite{Liu:2008zf}).
The main purpose of this work is to generalize the 
boundary counterterms and the quasilocal stress energy
tensor there to a positive value of the cosmological constant 
and thus to extend the prescription in  \cite{Balasubramanian:2001nb},
\cite{Ghezelbash:2001vs} to
the case of  EGB theory.
 Our results are valid for configurations with $d\leq 7$, although
 a general counterterm expression is also conjectured.
In the second part of this paper we apply this general formalism to several different
asymptotically  dS black holes. 
Apart from known static solutions, we consider also rotating black holes with two 
equal magnitude angular
momenta in  EGB theory formulated in
five spacetime dimensions. 
These solutions are constructed numerically within a nonperturbative approach, by directly solving the EGB equations
with suitable boundary conditions.
They also provide a nontrivial generalization in EGB theory of a  
 particular class of the known Myers-Perry-dS$_5$ black holes \cite{Gibbons:2004js}.

Most of the notation and sign conventions used in this paper are similar to
those in ref.~\cite{Ghezelbash:2001vs}.

%%%%%%%%%%%%%%%%%%%%%%%%%%%%%%%%%%%%%%%%%%%%%%%%%%%%%%%%%%%%%%%%%%
\section{The general formalism}
%%%%%%%%%%%%%%%%%%%%%%%%%%%%%%%%%%%%%%%%%%%%%%%%%%%%%%%%%%%%%%%%%%
%%%%%%%%%%%%%%%%%%%%%%%%%%%%%%%%%%%%%%%%%%%%%%%%%%%%%%%%%%%%%%%%%%
\subsection{The action and field equations}
%%%%%%%%%%%%%%%%%%%%%%%%%%%%%%%%%%%%%%%%%%%%%%%%%%%%%%%%%%%%%%%%%%
 We consider the EGB model with a positive cosmological constant $\Lambda= (d-2)(d-1)/2\ell^2$,
  coupled with 
  some matter fields  
 with a lagrangean density ${\cal L}_M$
\begin{eqnarray}
\label{action}
I=\frac{1}{16 \pi G}\int_\mathcal{M}~d^dx \sqrt{-g} \left(R-2 \Lambda+\frac{\alpha}{4}L_{GB} +{\cal L}_M\right),
\end{eqnarray}
where
% $R$ is the Ricci scalar  and
\begin{eqnarray}
\label{LGB}
L_{GB} =R^2-4R_{\mu \nu}R^{\mu \nu}+R_{\mu \nu \sigma \tau}R^{\mu \nu \sigma \tau},
\end{eqnarray}
is the GB term,
while $R$, $R_{\mu\nu}$  and $R_{\mu \nu \sigma \tau}$ are the Ricci scalar, the 
Ricci tensor and the Riemann tensor associated with the bulk metric $g_{\mu\nu}$. 
For $d=4$, $L_{GB}$ is a topological invariant and thus does not contribute to the equations of motion; in higher dimensions it is 
the most general quadratic expression which preserves the property that the equations of motion involve 
only second order derivatives of the metric. 
The constant $\alpha$ in (\ref{action}) is the GB coefficient with dimension $(length)^2$ and is positive
in the string theory.
We shall therefore restrict in this work to the case $\alpha>0$, although the counterterm expression does not depend on this choice.

The variation of the action (\ref{action}) with respect to the metric
tensor results in the gravity equations of the model
\begin{eqnarray}
\label{eqs}
R_{\mu \nu } -\frac{1}{2}Rg_{\mu \nu}+\Lambda g_{\mu \nu }+\frac{\alpha}{4}H_{\mu \nu}=2 T_{\mu \nu}~,
\end{eqnarray}
where
\begin{equation}
\label{Hmn}
H_{\mu \nu}=2(R_{\mu \sigma \kappa \tau }R_{\nu }^{\phantom{\nu}%
\sigma \kappa \tau }-2R_{\mu \rho \nu \sigma }R^{\rho \sigma }-2R_{\mu
\sigma }R_{\phantom{\sigma}\nu }^{\sigma }+RR_{\mu \nu })-\frac{1}{2}%
L_{GB}g_{\mu \nu }  ~,
\end{equation}
and $T_{\mu \nu} $ is the energy-momentum tensor of the matter fields.

For a well-defined variational principle, one has to supplement the 
action (\ref{action}) with the Gibbons-Hawking surface term \cite{GibbonsHawking1}
\begin{equation}
I_{b}^{(E)}=-\frac{1}{8\pi G}\int_{\partial \mathcal{M^-}}^{\partial \mathcal{M^+}}d^{d-1}x\sqrt{ \gamma }K~,
\label{Ib1}
\end{equation}
and its counterpart for the GB gravity  \cite{Myers:1987yn} 
\begin{equation}
I_{b}^{(GB)}=-\frac{\alpha}{16\pi G} \int_{\partial \mathcal{M^-}}^{\partial \mathcal{M^+}}d^{d-1}x\sqrt{ \gamma }%
 \left( s J-2{\rm G}_{ab} K^{ab}\right)  ~,
\label{Ib2}
\end{equation}
where 
$\gamma _{ab } $ is the induced metric on the boundary with the outward-pointing normal vector $n^a$  and  $K$ is  the
trace of the extrinsic curvature $K_{ab}$ of the boundary.
The factor $s$ in (\ref{Ib2})  is $s=+1$ for a spacelike normal vector $n^a$ and  $s=-1$ for a timelike normal vector  (the case considered here), 
see $e.g.$ \cite{Gravanis:2007ei}.
 Other quantities in (\ref{Ib2}) are ${\rm G}_{ab}$ -- the Einstein tensor of the metric $\gamma _{ab}$ and $J$ --the
trace of the tensor
\begin{equation}
J_{ab}=\frac{1}{3}%
(2KK_{ac}K_{b}^{c}+K_{cd}K^{cd}K_{ab}-2K_{ac}K^{cd}K_{db}-K^{2}K_{ab})~.
\label{Jab}
\end{equation}
Also, the case of interest in this paper corresponds to a spatial 
boundary future/past timelike infinity\footnote{For the black hole solutions in this paper, 
this corresponds to evaluate various quantities
for some fixed radius larger than the radius of the cosmological
horizon and then sending this radius to infinity.}.
Therefore ${\cal \partial M}^{\pm }$ are spatial Euclidean boundaries at early and late times, while
$\int_{{\cal \partial M}^{-}}^{{\cal \partial M}^{+}}d^{d-1}x$ indicates
an integral over the late time boundary minus an integral over the early
time boundary.
In what follows, to simplify the picture,
 we will consider the  $\mathcal{I}^{+}$ boundary only,
dropping the $^{\pm }$ indices (similar results hold for $\mathcal{I}^{-}$).

The equations (\ref{eqs}) present  many interesting solutions possessing new features
as compared to the pure Einstein gravity case (for a review, see the recent work 
\cite{Charmousis:2008kc}).
In this Section we shall consider the issue of computing the action  
and global charges of the EGB solutions in dS spacetime,
several examples of such configurations being discussed in the next Sections.
Here we have found  convenient to write\footnote{For the sake of simplicity, we have restricted
ourselves to the case of solutions with a well defined
Einstein gravity limit. However, the results in this section can easily be generalized to the branch of
solutions diverging as $\alpha\to 0$, in which case several terms in (\ref{Lagrangianct}) have an
opposite sign.  } 
\begin{eqnarray}
\label{lc}
\ell_c=\ell\sqrt{\frac{1+U}{2}},~~{\rm~~with~~~~}U=\sqrt{\frac{\alpha(d-3)(d-4)}{\ell^2}+1},
\end{eqnarray}
which results in a compact form for the expression below;
physically, $\ell_c$ corresponds to an effective dS length scale in EGB theory.

%%%%%%%%%%%%%%%%%%%%%%%%%%%%%%%%%%%%%%%%%%%%%%%%%%%%%%%%%%%%%%%%%%
\subsection{The counterterms and the boundary stress tensor}
%%%%%%%%%%%%%%%%%%%%%%%%%%%%%%%%%%%%%%%%%%%%%%%%%%%%%%%%%%%%%%%%%%
The GB term in (\ref{action}) does not change the 
 general formalism to compute the conserved charges and the action of 
asymptotically dS solutions developed  in \cite{Balasubramanian:2001nb},
\cite{Ghezelbash:2001vs}.
Therefore, we only recapitulate the basic steps here, emphasizing the new features 
which emerge for $\alpha\neq 0$.

In general, the action (\ref{action}) (together with the boundary terms (\ref{Ib1}), (\ref{Ib2}))   
is divergent when evaluated on a solution
to the equations of motion (\ref{eqs}).
In the counterterm approach, the remedy is to supplement the initial action (\ref{action}) 
by a 
 boundary counterterm part $I_{\mathrm{ct}}$
depending only on geometric invariants of the boundary metric (therefore the bulk equations of motion
remain the same). 
$I_{ct}$ regularizes the tree gravitational
action $I_{cl}$ and the boundary stress tensor.  
Crucial to the success of the counterterm prescription is that the divergencies are universal, so that a single choice
of the counterterms
suffices to render finite the action of all asymptotically dS solutions.

For $d<8$  solutions, our proposal for the boundary counterterm action is\footnote{Note that in 
odd spacetime dimensions, for some boundary geometries, there is an additional
logarithmic divergence that cannot be cancelled  without includding 
an explicit cutoff dependence
in the counterterm action, which should be supplemented with new extra terms.
This feature occurs already for Einstein gravity theory, leading
to a conformal anomaly similar to what has been obtained in the context of AdS spacetime \cite{Balasubramanian:1999re}. 
However, this is not the case of the solutions discussed in next sections.
}
\begin{eqnarray}
\label{Lagrangianct} 
I_{\mathrm{ct}} &=&\frac{1}{8\pi G}\int_{\partial \mathcal{M}} d^{d-1}x\sqrt{\gamma 
}
\bigg\{
  -(\frac{d-2}{\ell_c })(\frac{2+U}{3})
+\frac{\ell_c \mathsf{\Theta } \left( d-4  \right) } {2(d-3)}(2-U)\mathsf{R}
 \\
 \nonumber
 &&-\frac{\ell_c ^{3}\mathsf{\Theta }\left( d-6\right) }{2(d-3)^{2}(d-5)}
 \left[ U\bigg( \mathsf{R}_{ab}\mathsf{R}^{ab}-\frac{d-1}{4(d-2)}\mathsf{R}^{2} \bigg)
 -\frac{d-3}{2(d-4)}(U-1)L_{GB} 
  \right] 
\bigg\},
\end{eqnarray}
where $\mathsf{R}$, $\mathsf{R}^{ab}$  and $L_{GB}$ are the curvature, the 
Ricci tensor and the GB term associated with the induced metric $\gamma $. 
Also, $\mathsf{\Theta}(x)$ is the step-function with  $\mathsf{\Theta 
}\left( x\right) =1$ provided $x\geq 0$, and zero otherwise.
One can easily see that as $\alpha \to 0$ (thus $U \to 1$),  the known counterterm
expression in the Einstein gravity \cite{Balasubramanian:2001nb}, \cite{Ghezelbash:2001vs} is recovered.

Following ref. \cite{Brihaye:2008xu}, we conjecture the general
form of $I_{ct}$ in $d$ spacetime dimensions:
\begin{eqnarray}
\label{Ict-gen}
I_{\mathrm{ct}} &=&\frac{1}{8\pi G}\int_{\partial \mathcal{M}} d^{d-1}x\sqrt{ \gamma }
\bigg\{  
\sum_{k\geq 1} \mathsf{\Theta } \left( d-2k  \right)\bigg(f_1(U)L_E+f_2(U){\cal L}_{(k-1)} 
\bigg)
\bigg\} ,
\end{eqnarray}
where $L_E$ is the corresponding $k$-th part of
the counterterm lagrangean for a theory with only Einstein gravity in the bulk 
(with the length scale $\ell$ in front of it replaced
by the new effective dS radius $\ell_c$) and ${\cal L}_{(k-1)}$ is the $(k-1)$ term in the Lovelock hierarchy.
The functions $f_1(U)$, $f_2(U)$ are first order polynomials in $U$, whose expression can easily be derived from
those given in \cite{Brihaye:2008xu} for $\Lambda<0$. 
The series (\ref{Ict-gen})
truncates for any fixed dimension, with new terms entering at every 
new even value of $d$.
 
Once we know the  expression of the boundary counterterm, the computation of the conserved charges is performed in 
 a similar way to the $\alpha=0$ limit \cite{Balasubramanian:2001nb,Ghezelbash:2001vs}.
The (Euclidean) boundary metric on equal time surfaces can be written, at least locally, in a ADM-like general form 
\begin{equation}
ds^{  2}=\gamma_{ab}^{  }d {x}^{ a}d {x}^{ b}=N_{\rho}^{ 2}d\rho ^{2}+\sigma _{ab}^{ }\left( d\psi ^{  a}+N^{  a}d\rho
\right) \left( d\psi ^{ b}+N^{  b}d\rho \right) ,  
\label{hmetric}
\end{equation}%
where $N_{\rho }$ and $N^{a}$ are the lapse function and the shift vector
respectively, while $\psi ^{a}$ are angular variables parametrizing  a closed
surfaces $\Sigma$.  
The physical significance of the coordinate $\rho$ in (\ref{hmetric}) depends on the 
considered situation; $e.g.$
for the black hole solutions discussed in the next Sections, $\rho$
is the coordinate associated with the asymptotic Killing vector
that is timelike inside the static patch of dS, but spacelike at $\mathcal{I}^{\pm}$.

Varying the total action with respect to the boundary metric $\gamma _{ab}$ 
results in the following  boundary stress-energy tensor
\begin{equation}
\mathsf{T_{ab}}=\frac{2}{\sqrt{ \gamma }}\frac{\delta }{\delta \gamma^{ab}}\left(
I+I_{b}^{(E)}+I_{b}^{(GB)}+I_{\text{ct}}  \right) ,
\label{Tab}
\end{equation}
with the following expression valid for $d<8$:
%%%%%%%%%%%%%%%%%%%%%%%%%%%%%%%%%%���
\begin{eqnarray}
8 \pi G \mathsf{T_{ab}}=
K_{ab}-\gamma _{ab}K
 +\frac{{\alpha}}{2} (Q_{ab}-\frac{1}{3}Q\gamma_{ab}) 
+\frac{d-2}{\ell_c}\gamma _{ab}(\frac{2+U}{3})
+\frac{\ell_c
\mathsf{\Theta } \left( d-4  \right) }{d-3}(2-U)%
\left( \mathsf{R}_{ab}-\frac{1}{2}\gamma _{ab}\mathsf{R} \right)  
\notag 
\\
 +\ell_c^{3}\mathsf{\Theta } \left( d-6  \right) 
\bigg\{
-\frac{U}{(d-3)^{2}(d-5)}
 \bigg(
   -\frac{1}{2}\gamma _{ab}\left(
\mathsf{R}_{cd}\mathsf{R}^{cd}-\frac{(d-1)}{4(d-2)}\mathsf{R}^{2}\right)  
-\frac{(d-1)}{2(d-2)}%
\mathsf{RR}_{ab} 
+2\mathsf{R}^{cd}\mathsf{R}_{cadb}
\notag 
\\
\label{TabCFT}  
-\frac{d-3}{2(d-2)}\nabla _{a}\nabla _{b}\mathsf{R}+\nabla
^{2}\mathsf{R}_{ab}-\frac{1}{2(d-2)}\gamma _{ab}\nabla ^{2}\mathsf{R } 
\bigg )
 +\frac{ U-1 }{2(d-3) (d-4)(d-5)}
\mathsf{H}_{ab}
\bigg \}+\dots
\end{eqnarray}
where  \cite{Gravanis:2007ei}, \cite{Davis:2002gn}
\begin{eqnarray}
&Q_{ab}= 
s\left ( 2KK_{ac}K^c_b-2 K_{ac}K^{cd}K_{db}+K_{ab}(K_{cd}K^{cd}-K^2) \right)
+2K \mathsf{R}_{ab}+\mathsf{R}K_{ab}
-2K^{cd}\mathsf{ R}_{cadb}-4 \mathsf{R}_{ac}K^c_b,~{~~~~~~}
\end{eqnarray}
and $\mathsf{H}_{ab}$ given by (\ref{Hmn}), this time written in terms of the boundary metric $\gamma_{ab}$, however.
 All terms in (\ref{TabCFT}), except the first four,
 come from the variation of the counterterms in (\ref{Lagrangianct}).
The boundary stress-energy tensor $\mathsf{T_{ab}}$ 
measure the response of the spacetime 
to changes of the boundary metric and encodes the notion of conserved global charges.

Following \cite{Balasubramanian:2001nb},
\cite{Ghezelbash:2001vs}, let us suppose that $\xi^{ i }$ is a  Killing vector generating an isometry of 
the boundary geometry (\ref{hmetric}).
Then it is straightforward to show that $\mathsf{T_{ij}}^{  }\xi ^{  j}$ is divergenceless and
one can define a  conserved  quantity $\mathfrak{Q}_{\xi }$ associated with $\xi^{ i }$
as follows
\begin{equation}
\label{Qcons}
\mathfrak{Q_{\xi}}{}^{  }=\oint_{\Sigma  }d^{n}\psi ^{  }\sqrt{%
\sigma ^{  }}n^{  i}\mathsf{T_{ij}}^{  }\xi ^{  j} ,  
\end{equation}%
where $n^{ i}$ is 
a unit vector normal on a surface of fixed $\rho$.
The physical interpretation of this relation is the same for any theory of gravity:
it means that a collection of
observers, on the hypersurface with the induced metric $\sigma_{ij}$, would all
measure the same value of $\mathfrak{Q}_{\xi }$ provided this surface has an
isometry generated by $\xi ^{i}$.  
As mentioned above, a dS spacetime has no globally timelike
Killing vector, which makes difficult
to define a mass for the solutions with this asymptotics.
However, for all cases of interest ($e.g.$ the black holes solutions
in the next Sections), there is a Killing vector that is timelike inside a
 static patch, while it is spacelike outside the cosmological horizon and therefore
 at $\mathcal{I}^{\pm}$.
(Moreover, any spacetime that is asymptotically dS will have such an asymptotic symmetry generator.)
 The total mass/energy of solutions is evaluated with respect to this Killing vector.
 
Proceeding further, one can define a Hawking temperature $T_H$ 
($e.g.$ by computing the corresponding surface gravity) and entropy $S$ for the cosmological 
horizon by  using the
saddle point approximation to the gravitational partition function 
(namely the generating functional analytically 
continued to the Euclidean spacetime). 
In the semiclassical approximation, the dominant 
 contribution to the path integral will 
come from the neighborhood of saddle points of the action, that is, of classical 
solution; the zeroth order contribution to 
$\log Z$ is given by $-I_{cl}$.
A tree-level evaluation of the path integral with a GB term may be carried out along the lines
described $e.g.$ in ref. \cite{Ghezelbash:2001vs} for the Einstein gravity case. 
Therefore, we find the entropy of the cosmological horizon (with $\beta=1/T_H$)
\begin{equation}  
\label{S}
S=\beta (E-\mu _{i}{\frak C}_{i})-I_{cl},  
\end{equation} 
which is found upon application of the Gibbs-Duhem relation to the partition 
function, with chemical potentials ${\frak C}_{i}$ and
conserved charges $\mu _{i}$,  while $E$ is the total mass/energy, evaluated according to (\ref{Qcons}).
Also, all solutions should satisfy the
 first law of thermodynamics  for the cosmological horizon
\begin{equation}
dS=\beta (dE-\mu _{i}d{\frak C}_{i}), 
\label{1stlaw}
\end{equation}  
 which provides a test of the general formalism.

%%%%%%%%%%%%%%%%%%%%%%%%%%%%%%%%%%%%%%%%%%%%%%%%%%%%%%%%%%%%%%%%%%
\section{Applications: known solutions}
%%%%%%%%%%%%%%%%%%%%%%%%%%%%%%%%%%%%%%%%%%%%%%%%%%%%%%%%%%%%%%%%%%
%%%%%%%%%%%%%%%%%%%%%%%%%%%%%%%%%%%%%%%%%%%%%%%%%%%%%%%%%%%%%%%%%%
\subsection{dS spacetime in EGB theory}
%%%%%%%%%%%%%%%%%%%%%%%%%%%%%%%%%%%%%%%%%%%%%%%%%%%%%%%%%%%%%%%%%%
As the simplest illustration of the above formalism, we consider the case of  
empty dS spacetime.
This solution has a simple form in a
large number of coordinate systems.
For example, there is a static frame centered on each
observer (timelike geodesic) in dS. Moreover, when a black hole exists, there is still
a static frame centered about the black hole. Since different
parametrizations emphasize different features, it is of interest to consider
dS spacetime in alternative coordinate systems corresponding to different
classes of observers.

Starting with an inflationary coordinate system, the dS solution reads
\begin{eqnarray}
\label{inf}
ds^{2} =-dt^2+e^{2t/\ell_c}d\vec x^2~,
\end{eqnarray}
which solves the EGB equations (\ref{eqs}) with $T_{\mu\nu}=0$ ($i.e.$ no matter fields).
The properties of this solution are similar to the case of Einstein gravity (see $e.g.$ \cite{Klemm:2004mb});
the equal time surfaces here are flat, while $t$ runs from $-\infty$ to $+\infty$.
One can easily verify that the counterterms  (\ref{Lagrangianct}) removes
all divergencies of the total action for $d\leq 7$, and leads to $I_{cl}=0$.
The total mass/energy of this solution is also vanishing, since $\mathsf{T_{ij}}=0$.

The situation is different for a static coordinate system,
the corresponding line element being
\begin{eqnarray}
ds^{2} =\frac{dr^{2}}{F(r)}+r^{2}d\Omega _{d-2}^{2}-F(r)dt^{2},  
\label{dS}
\end{eqnarray}
where (here we shall consider only the branch of solutions
with a smooth Einstein gravity limit)
\begin{eqnarray}
F(r) = 1+\frac{2r^2}{\alpha (d-3)(d-4)}
 \bigg (
 1-\sqrt{1+\alpha (d-3)(d-4)\frac{1}{\ell^2})}
 \bigg)=1-\frac{r^2}{\ell_c^2}.
\nonumber
\end{eqnarray}
This spacetime has a cosmological horizon at $r_c=\ell_c$ (where $F(r_c)=0$), with an associated temperature $T_H^c=(2\pi \ell_c)^{-1}$.
The topology of this solution for large constant $r>r_c$, is an Euclidean cylinder $R\times S^{d-2}$
and $t$ is the coordinate along the cylinder. $\mathcal{I}^{\pm }$ are
located outside the future/past cosmological horizons, where $r$ is timelike
and $t$ is spacelike.
The relationship between the coordinate patches 
(\ref{inf}) and (\ref{dS}) and their Penrose diagrams are presented in ref. \cite{Hawking}.

The general formalism in Section 2 is applied 
 working
outside of the cosmological horizon, where $F(r)<0$. 
The gravitational mass/energy is the charge associated with the Killing
vector $\partial /\partial t$ --- now spacelike outside the cosmological
horizon. As expected, the total energy found by using the counterterm prescription vanishes
for an even dimensional spacetime and has a nonzero value for an odd $d$:
\begin{eqnarray}
\label{Casimir}
M_0={V_{d-2}\over 8\pi G} {(d-2)!!^2\over (d-1)!} 
\left(\frac{(d-2)U-2}{d-4} \right)\ell_c^{d-3} \delta_{2p+1,d}~~,
\end{eqnarray}% 
where $V_{d-2}$ is the area of the unit $S^{d-2}$ sphere and $p\geq 2$ is an integer.
For solutions in Einstein gravity ($\alpha=0$), $M_0$ is usually interpreted
as the Casimir energy in the context of dS/CFT correspondence.
Also, it reduces to the expression
obtained in ref.~\cite{Ghezelbash:2001vs} when $U=1$.

From (\ref{S}) one finds the following expression for the entropy of dS spacetime in EGB theory: 
%(note the nontrivial dependence on $\alpha$): 
\begin{eqnarray}
\label{dS-entropy1}
 S= \frac{V_{d-2}}{4G}\ell_c^{d-4}\left (\ell_c^2+\frac{\alpha}{2}(d-2)(d-3) \right),
\end{eqnarray}  
which in the limit of small $\alpha$ can written in the simple form
\begin{eqnarray}
\label{dS-entropy2}
 S=S_0+S_{c}~~~{\rm with}~~~S_0=\frac{\ell^{d-2}V_{d-2}}{4G}~,
~~~~S_c=\alpha  \frac{V_{d-2}}{4G}d(d-2)(d-3) \ell^{d-4}.
\end{eqnarray}

From the study of (\ref{inf}), (\ref{dS}) we conclude that, similar to the case of Einstein gravity,
the horizon and entropy of the dS space in EGB theory have an obvious observer
dependence.  

%%%%%%%%%%%%%%%%%%%%%%%%%%%%%%%%%%%%%%%%%%%%%%%%%%%%%%%%%%%%%%%%%%
\subsection{Reissner--Nordstr\"om--dS-GB black hole }
%%%%%%%%%%%%%%%%%%%%%%%%%%%%%%%%%%%%%%%%%%%%%%%%%%%%%%%%%%%%%%%%%%
These solutions are found for a matter lagrangean density ${\cal L}_M=-{\cal F}^2$, with the Maxwell
field strength tensor ${\cal F}=dA$, 
where the (pure electric-) gauge
potential is 
\begin{eqnarray}  
\label{A}
A=A_tdt=\Big(\sqrt{\frac{d-2}{2(d-3)}}\frac{ Q}{r^{d-3}}+\Phi \Big) dt,
\end{eqnarray}
where $\Phi$ is a constant. 
Working again in a static coordinate system,
the line element is still given by (\ref{dS}), with a different expression for $F(r)$, however:
\begin{eqnarray}  
\label{F}
F(r)=   1+\frac{2r^2}{\alpha (d-3)(d-4)}
 \bigg (
 1-\sqrt{1+\alpha (d-3)(d-4)(\frac{M}{r^{d-1}}-\frac{Q^2}{r^{2(d-2)}}+\frac{1}{\ell^2})}
 \bigg).
\end{eqnarray}
As argued below, $M$
and $Q$  in the above expression  are constants proportional to the gravitational mass~/energy $E$
and the total electric charge $\mathbf{Q}$, respectively. 
The $Q=0$ limit of this metric
corresponds to the EGB generalization of the McVittie solution describing a Schwarzschild
black hole embedded in dS spacetime \cite{Shiromizu:2001bg}.

A discussion of the solution (\ref{A}),  (\ref{F}) appeared in ref.~\cite{Torii:2005xu} 
(see also  ref. \cite{Astefanesei:2003gw} for an extended analysis 
of  the limiting case $\alpha=0$, including also multi-black hole generalizations). 
Here we briefly review its basic properties.
One can easily verify that the metric has a curvature singularity at the origin $r=0$. 
In general,
the metric (\ref{dS}) presents Killing horizons at the radii where $F(r)$ vanishes.
Of interest are the outer black hole horizon at $r=r_{h}$ and the
cosmological horizon $r=r_{c}$ corresponding to the largest root of $F(r)$. 
The Hawking temperature of associated to each of the horizons  is $T_H^{h,c}=|F'(r_{h,c})|/(4 \pi)$, where a prime denotes 
the derivative with respect the radial coordinate.
The two horizons are not in thermal equilibrium because the time
periods in the Euclidean section required to avoid a conical singularity at
both do not match in general. 
An extremal black hole is found by imposing $F(r_h)=F'(r_h)=0$
which fixes $M$, $Q$ as functions of $\ell,\alpha$ and $r_h$ 
(a similar relation is found when considering instead the cosmological horizon).
The constant $\Phi$ in (\ref{A}) is usually fixed such that $A_t(r_c)=0$, 
and thus it corresponds to the electrostatic difference between
the cosmological horizon and infinity. 

The computation of the mass, action and entropy of a RNdS black hole is a
direct application of the method described in the previous section. 
The gravitational mass/energy is the charge associated with the Killing
vector $\partial /\partial t$. 
The total mass/energy found by using the counterterm prescription described in the previous
Section is 
\begin{equation}
E=-\frac{V_{d-2}}{16\pi G}(d-2)M +M_0,
\label{energy}
\end{equation}  
with $M_0$ the Casimir term given by (\ref{Casimir}).
The negative sign implies that the black hole lowers the total bulk
energy with respect to the total energy of the pure dS spacetime \cite{Balasubramanian:2001nb}.

The computation of the total electric charge is similar to that performed
in \cite{Astefanesei:2003gw} for the  case without a GB term.  The results there show that 
 the total electric charge evaluated at future/past infinity is
\begin{eqnarray}
\mathbf{Q}= \frac{QV_{d-2}}{8 \pi G}\sqrt{ 2(d-3)(d-2)}.
\end{eqnarray}
From (\ref{S}) one finds the  entropy of the comological horizon 
(note that both $S_0$ and $S_c$ have a  nontrivial dependence on $\alpha$): 
\begin{eqnarray}
\label{BH-entropy}
 S=S_0+S_{c}~~~{\rm with}~~~S_0=\frac{V_{ d-2}}{4G}r_c^{d-2}~,
~~~~S_c=\alpha  \frac{V_{ d-2}}{4G}\frac{1}{2} r_c^{d-4}(d-2)(d-3).
\end{eqnarray}  
%with  the event horizon area $A_c=r_c^{d-2}V_{ d-2}$.
One can easily verify that the first law of thermodynamics (\ref{1stlaw}) also holds, with $\mu _{i}=\Phi,~~{\frak C}_{i}=\mathbf{Q}$.

It would be interesting to study the properties of
the  Reissner--Nordstr\"om--de Sitter solution (\ref{A}), (\ref{F}) in the
 inflationary coordinate system (\ref{inf}).
This problem has been considered in  ref. \cite{Astefanesei:2003gw} in the absence of a GB term in the action.
Interestingly, the same general picture have been found there for black holes in 
both coordinate systems (\ref{inf}) and (\ref{dS}),
which shows the complex relation between different classes of observers in
dS spacetime.
For example, the mass of the 
 black holes in an inflationary coordinate system is still given by (\ref{energy}), 
 with $M_0=0$ however\footnote{This result has been interpreted in \cite{Astefanesei:2003gw}
as providing support for the putative dS/CFT
correspondence, since the general features of the CFT dual to a black
hole should not depend on the dS slicing choice.}.
We expect that a similar result will be found in the presence of a GB term.
However, in the absence of an explicit form of the  Reissner--Nordstr\"om--dS-GB black hole
 in the inflationary coordinate 
 system{\footnote{The main obstacle here is the absence  of a simple closed form expression 
of the Reissner--Nordstr\"om--GB (or even Schwarzschild-GB) black hole in an isotropic 
coordinate system for
the $\Lambda=0$ case. For $\alpha=0$, this form of the solution is used
to construct cosmological configurations by using the prescription in \cite{Kastor:1992nn}.}, 
any progress in this direction appears to require a separate numerical study of these solutions.
 
%%%%%%%%%%%%%%%%%%%%%%%%%%%%%%%%%%%%%%%%%%%%%%%%%%%%%%%%%%%%%%%%%%
\section{Rotating EGB black holes with positive cosmological constant}
%%%%%%%%%%%%%%%%%%%%%%%%%%%%%%%%%%%%%%%%%%%%%%%%%%%%%%%%%%%%%%%%%
%%%%%%%%%%%%%%%%%%%%%%%%%%%%%%%%%%%%%%%%%%%%%%%%%%%%%%%%%%%%%%%%%%
\subsection{The metric ansatz and known limits}
%%%%%%%%%%%%%%%%%%%%%%%%%%%%%%%%%%%%%%%%%%%%%%%%%%%%%%%%%%%%%%%%%%
 The computation of the global charges and entropy of a
rotating black holes in EGB theory represents another nontrivial aplication
of the general formalism in Section 2.
Unfortunately,  no exact solutions are available in this case, and one has to 
solve  numerically the field equations.

To simplify the general picture, we consider here the vacuum case in $d=5$ dimensions  only, 
although
the inclusion of a U(1) field is straightforward in principle.
A general  spinning black hole solution is characterized in this case
by two angular momenta and its mass/energy,  and can be found by solving
a set of seven partial differential equations.
 However,  the numerical problem
is greatly simplified by taking the ${\it a priori}$ independent 
two angular momenta to be equal in order to factorize the 
angular dependence \cite{Kunz:2005nm}, \cite{Kunz:2006eh}. 
The asymptotic expressions and the explicit computation of 
the action and boundary stress tensor also simplifies drastically in this case. 

To construct these solutions, we use the same metric ansatz employed in ref. \cite{Brihaye:2008kh} for 
the corresponding problem with $\Lambda<0$:
\begin{eqnarray}
\label{metric}
&&ds^2 = \frac{dr^2}{f(r)}
  + g(r) d\theta^2
+h(r)\sin^2\theta \left( d \varphi -w(r)dt \right)^2 
+h(r)\cos^2\theta \left( d \psi -w(r)dt \right)^2 
\\
\nonumber
&&{~~~~~~}+(g(r)-h(r))\sin^2\theta \cos^2\theta(d \varphi -d \psi)^2
-b(r) dt^2~,
\end{eqnarray}
where $\theta  \in [0,\pi/2]$, $(\varphi,\psi) \in [0,2\pi]$, $r$ and $t$ being the
radial and time coordinates.  
This ansatz has a residual degree of freedom which is fixed by taking $g(r)=r^2$.

The equations satisfied by the functions $b,f,h,w$ result  directly from (\ref{eqs}).
We refrain to write them because they are very long and not particularly enlightening.
They present however two exact solutions which are important in what follows.
 The dS$_5$ generalization \cite{Gibbons:2004js} of the  Myers-Perry rotating  
 black holes \cite{Myers:1986un}  with equal magnitude angular momenta (hereafter MPdS$_5$)
is found for $\alpha=0$ (no GB term) 	and has
\begin{eqnarray}
\label{MP-AdS}
f(r)=1-\frac{r^2}{\ell^2}
-\frac{2{\hat M}\Xi}{r^{2}}
+\frac{2{\hat M}{\hat a}^2}{r^{4}},~
h(r)=r^2\left(1+\frac{2{\hat M}{\hat a}^2}{r^{4}}\right),~
%\\
%\nonumber
w(r)=\frac{2{\hat M}{\hat a}}{r^2 h(r)},~
g(r)=r^2,~~ b(r)=\frac{r^2 f(r)}{h(r)},
\end{eqnarray}
where ${\hat M}$ and ${\hat a}$ are two constants related to the solution's mass and 
angular momentum, while $\Xi=1+{\hat a}^2/\ell^2$.

For $g(r)=h(r)=r^2,~~w(r)=0$ and $f(r)= b(r)=  1+  r^2/{\alpha  }
 \bigg (
 1-\sqrt{1+2 \alpha ( {M}/{r^{4}} + {1}/{\ell^2})}
 \bigg)$, one recovers the Schwarzschild-dS$_5$ solution with a Gauss-Bonnet term. 
The slowly rotating generalisation of this solution\footnote{By using the results derived in \cite{Kim:2007iw}
for $\Lambda<0$, one can find a different set of  closed form EGB
asymptotically dS$_d$ rotating black hole solutions with only one nonvanishing angular momentum 
(where the rotation parameter appears as a small quantity), 
the effects of an U(1) field being also included.
A different approximation of the rotating black hole solution
of the $d=5$ EGB equations with a cosmological constant 
have been presented in closed form in \cite{Alexeyev:2007sd}.} is found for 
  small values of the rotation parameter $a$, and reads
\begin{equation}
\label{w-slow}
w(r)=\frac{2a U^2}{\ell_c^2(U-1)}\left( \sqrt{1+\frac{2M \ell_c^2 (U-1)}{r^4U^2}}-1 \right ),
\end{equation} 
the other metric function remaining unchanged to this order in $a$.

%%%%%%%%%%%%%%%%%%%%%%%%%%%%%%%%%%%%%%%%%%%%%%%%%%%%%%%%%%%%%%%%%%
\subsection{Boundary conditions and global charges}
%%%%%%%%%%%%%%%%%%%%%%%%%%%%%%%%%%%%%%%%%%%%%%%%%%%%%%%%%%%%%%%%%%

We want the generic line element (\ref{metric}) to describe a nonsingular,
asymptotically de Sitter spacetime outside a cosmological horizon located at $r=r_c>0$, with $f(r_c)=0$.
Here $f(r_c)=0$ is only a coordinate singularity. The regularity assumption implies that all
curvature invariants at $r=r_c$ are finite.
 Outside the cosmological horizon $r$ and $t$ changes the character
($i.e.$  $r$ becomes a timelike  coordinate  for $r>r_c$).
A nonsingular extension across this null
surface can be found just as at the event horizon of a black hole.
These configurations possess also an event horizon located at 
some intermediate value of the radial coordinate
$0<r_h<r_c$, all curvature
invariants being also finite as $r \to r_h$.  

 Restricting  to
nonextremal solutions, the following expansion holds near the event horizon 
 with the parameters $f_1^h,~b_1^h$,~$w_h^h$ and $h_h^h$, where $(f_1^h,~b_1^h,~h_h^h)>0$:
\begin{eqnarray}
\label{c1}
&&f(r)=f_1^h(r-r_h)+  O(r-r_h)^2,~~h(r)=h_h^h+ O(r-r_h),
 \\
 \nonumber
&&b(r)=b_1^h(r-r_h)+O(r-r_h)^2,~~w(r)=w_h^h+ O(r-r_h).~{~ }
\end{eqnarray}
A similar expansion holds for cosmological horizon, the corresponding parameters there
being $f_1^c,~b_1^c,~h_h^c$, and $w_h^c$.

Both the event  and the cosmological
horizon have their own  surface gravity $\kappa^{h,c}$, the associated Hawking temperatures being 
\begin{eqnarray} 
\label{Temp-rot} 
  T_H^{h,c}=\frac{|\kappa^{h,c}|}{2\pi}=\frac{\sqrt{ b_1^{h,c}f_1^{h,c}}}{4\pi}.
\end{eqnarray} 
Another quantities of interest are
the area $A_H^{h,c}$ of the black hole/cosmological horizon 
\begin{eqnarray}
\label{A2} 
A_H^{h,c}= \sqrt{h_h}^{h,c}  (r_h^{h,c})^2 V_3,
\end{eqnarray} 
where $V_3=2\pi^2$ denotes the area of the unit three dimensional sphere.

The Killing vector  $\chi=\partial/\partial_t+
 \Omega_\varphi \partial/\partial \varphi + \Omega_\psi \partial/\partial \psi $ is 
orthogonal to and null on both horizons. For the solutions 
within the ansatz (\ref{metric}), the 
event horizon's angular velocities are 
all equal, $\Omega_\psi^{h,c}=\Omega_\varphi^{h,c}=\Omega_H^{h,c}=w(r)|_{r=r_{h,c}} $.

A direct computation reveals that the solution 
admits at large $r$  a power series expansion of the form:
\begin{eqnarray}
\label{exp_inf}
 &&f(r)= 1-\frac{r^2}{\ell_c^2}+\sum_{k\geq 1}f_{2k}\left(\frac{\ell_c}{r}\right)^{2k},~~~
 b(r)= 1-\frac{r^2}{\ell_c^2}+\sum_{k\geq 1}b_{2k}\left(\frac{\ell_c}{r}\right)^{2k},~~~
 \\
 \nonumber
 &&h(r)= r^2\bigg(1+\sum_{k\geq 1}h_{2k}\left(\frac{\ell_c}{r}\right)^{2k}\bigg),~~~
w(r)=\frac{1}{r}\sum_{k\geq 1}w_{2k+1}\left(\frac{\ell_c}{r}\right)^{2k+1},
\end{eqnarray}
 where the coefficients $f_{2k},b_{2k},h_{2k},w_{2k+1}$ with $k>1$ are determined
 by $f_2,b_2$ and $w_3$. Specifically, we find $f_4=h_4=b_2-f_2,~b_6=(b_2(f_2+b_2(U-2))+(3U-2)w_3^2)/(2U),~w_7=-(f_2+U(2b_2-3f_2))w_3/(2U)$,
for the lowest order nonvanishing terms. Their expression becomes more complicated for higher $k$, with no general
pattern becoming apparent.
  
 The mass/energy\footnote{In the expression of $E$, we have subtracted 
 the Casimir energy of the pure dS$_5$ space as given by (\ref{Casimir}).} $E$
 and angular momenta of  these solutions evaluated at future/past timelike infinity
 by using the
counterterm formalism are fixed by the constants $f_2$, $b_2$ and $w_3$,  and read
\begin{equation}
                     E = \frac{V_3 }{16 \pi G}  \ell_c^2 U(4 b_2- f_2 )  \ \ , \ \ 
                     J_\varphi=J_\psi=J= -\frac{V_3 }{8 \pi G}\ell_c^3U w_3~.
\end{equation} 
The entropy of these solutions associated with the cosmological horizon is found from the relation (\ref{1stlaw})
with $\mu _{i}= \Omega_{\psi,\phi}^{c},~{\frak C}=J$:
\begin{eqnarray}
S=S_0+S_{GB},~~{\rm with}~~S_0=\frac{A_H^c}{4G} ,~~
S_{GB}=\alpha\frac{ V_3}{4G}\sqrt{h_h^c}(4-\frac{h_h^c}{(r_h^c)^2}).
\end{eqnarray}

%%%%%%%%%%%%%%%%%%%%%%%%%%%%%%%%%%%%%%%%%%%%%%%%%%%%%%%%%%%%%%%%%%
\subsection{The numerical method}
%%%%%%%%%%%%%%%%%%%%%%%%%%%%%%%%%%%%%%%%%%%%%%%%%%%%%%%%%%%%%%%%%%
Finding numerical solutions of a field theory model in a dS spacetime is a notoriously difficult task.
Therefore, before describing the properties of the solutions,
we shall give some details on the numerical methods we have used\footnote{The our knowledge, this is
the first attempt in the literature to   numerically  construct
EGB rotating solutions in a dS background. The approach and the numerical
methods here are quite different from those employed $e.g.$ in \cite{Brihaye:2008kh}
for rotating EGB solutions with AdS asymptotics or for
rotating Einstein-Maxwell black hole solutions  \cite{Kunz:2005nm}, \cite{Kunz:2006eh}.}.
The EGB field equations were solve
by employing a collocation
 method for 
boundary-value ordinary
differential equations, equipped with an adaptive mesh selection procedure
\cite{COLSYS}.
Typical mesh sizes include $10^3-10^4$ points.
The solutions have a typical relative accuracy of $10^{-8}$. 
In constructing rotating EGB-dS black holes, 
we make use of the existence of the  MPdS$_5$ and Schwarzschild-GB-dS closed form solutions, 
and employ them as starting configurations,
increasing  gradually $\Omega_H^{h}$ or $\alpha$, respectively.

However, when trying to find black hole solutions with $\Lambda>0$ for $r \in [r_h,\infty]$
 by imposing a regular horizon  at $r=r_h$, one has to tackle the technical difficulty that 
 there also appear a cosmological horizon\footnote{We do not consider in this work the behaviour of solutions
 inside the black hole even horizon $r<r_h$.}. That is, the metric functions
 $f,b$ admit a zero at an intermediate value of the variable $r$, say at $r=r_c>r_h$. 
 Of course, the value of $r_c$ is not known {\it apriori} as a function of $\Lambda,\alpha$.
  However, the condition 
 of a regular horizon should be imposed both at $r=r_h$ and $r=r_c$.
%The best way we could find to solve the equations numerically is to 
In our approach, we impose by hand the
values of $r_h,r_c$ and solve the equations first for $r\in [r_h,r_c]$ as a boundary value problem.
At the same time, we compute the value of $\Lambda$ corresponding to this cosmological horizon
by using the fictious equation $d\Lambda/dr=0$.
In a second step, we finally integrate 
the equations for $r\in [r_c,\infty]$ as an initial value problem with this value of the cosmological constant.
This assures that the metric functions and their first and second derivatives are continuous at $r=r_c$.

 In this approach, the set of boundary condition we imposed at $r_h,r_c$ is
 \begin{eqnarray}
 f=0,~~ b = 0,~~b' = 1,~~G(g_{ij},g'_{ij}) = 0,~~w= w_h  \ \ {\rm at} \ \ r = r_h,\\
  {\rm and}~~~~
 f=0~~ ,~~  b = 0,~~ G(g_{ij},g'_{ij}) = 0 \ \ {\rm at} \ \ r = r_c,
 \end{eqnarray}
 where $G(g_{ij},g'_{ij})$ is a complicated expression in terms of the metric function and their first derivatives which occurs from the
 condition for a regular horizon. 
In the above expression, the arbitrary rescaling of time is used to set $b'(r_h)=1$, keeping in mind that the
 function $b(r),~w(r)$ have to be renormalized at the end of the second step 
 according to
 \beq
          b(r) \to \tilde b(r) = b(r) \mu^2 \ \ , \ \ w(r) \to \tilde w(r) = w(r) \mu,
 \eeq
 where the constant $\mu$ is chosen in  such a way that the
 $\tilde b(r)$ approaches the asymptotic (\ref{exp_inf}).
  
 One disadvantage of this method is that the solutions cannot be studied systematically for fixed $\Lambda$.
  For the same reason, we have found difficult to study families of
 solutions obtained by varying $\alpha$ while $\Omega_H^{h}$ is fixed.
 
 To summarise, in our approach the input parameters are the black hole event horizon radius $r_h$, 
 the cosmological horizon radius $r_c$, the black hole  event horizon velocity $\Omega_H^{h}$ and the GB
 coupling parameter $\alpha$.
 The value  of the cosmological constant, the metric functions and their derivatives at $r=r_c$ 
 and the global charges emerge from the numerical output.

   %%%%%%%%%%%%%%%%%%%%%%%%%%%%%%%%%%%%
\begin{figure}[ht]
\hbox to\linewidth{\hss%
	\resizebox{11cm}{8cm}{\includegraphics{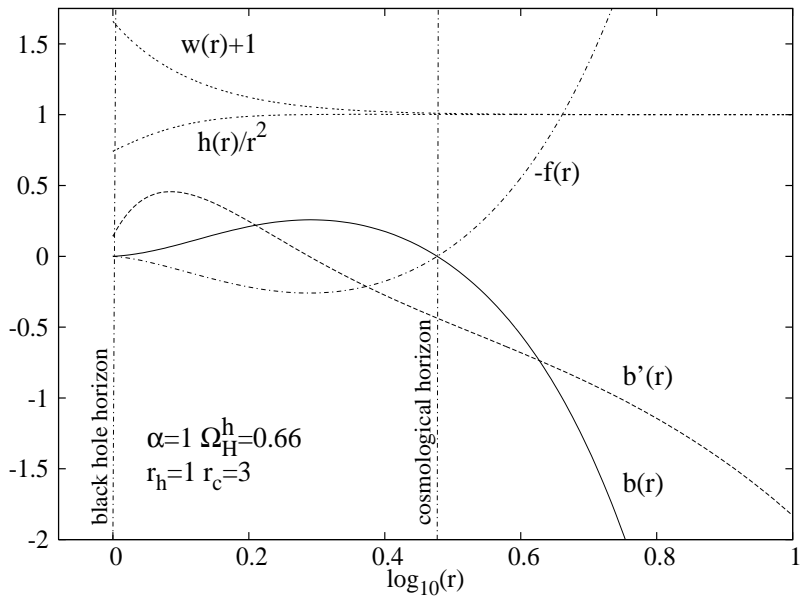}}
\hss}
 	\caption{ The profiles for a generic $d=5$ rotating black hole solution in EGB theory with positive cosmological constant.}
\label{Fig1}
\end{figure}
%%%%%%%%%%%%%%%%%%%%%%%%%%%%%%%%%%%%

%%%%%%%%%%%%%%%%%%%%%%%%%%%%
\subsection{Numerical results}
%%%%%%%%%%%%%%%%%%%%%%%%%%%
A systematic study of the properties of these rotating black 
holes appears to be a difficult task and
is beyond the purposes of this work.
In practice we have solved the equations numerically for several values of $r_h$, $r_c$ and $\Omega_H^{h}$ and a range of the 
Gauss-Bonnet coupling constant $\alpha$.

 %%%%%%%%%%%%%%%%%%%%%%%%%%%%%%%%%%%%
\begin{figure}[ht]
\hbox to\linewidth{\hss%
	\resizebox{11cm}{8cm}{\includegraphics{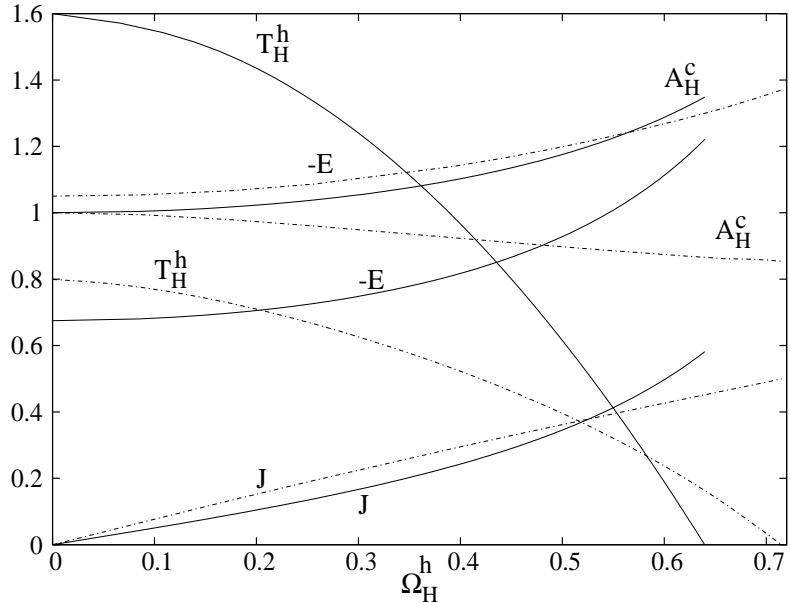}}
\hss}
 	\caption{  Several parameters are compared for the Myers-Perry-dS$_5$ (dotted curves) 
 	and Einstein-Gauss-Bonnet  (continuous line)
rotating black hole solutions.
The input parameters here were $r_h=1$, $r_c=3$, $\alpha=0.5$ and $\ell^2=10$. }
\label{Fig2}
\end{figure}
 %%%%%%%%%%%%%%%%%%%%%%%%%%%%%%%%%%%% 
 
 When increasing from zero the angular velocity $\Omega_H^{h}$, we have found numerical evidence for the existence of 
 nontrivial generalizations of any static Schwarzschild-GBdS configuration we considered;
the shape of the metric function $w(r)$ we found for small values of $\Omega_H^{h}$ is in good agreement with
(\ref{w-slow}).
We reach the same conclusion when considering instead GB counterparts of the Einstein gravity rotating solution (\ref{MP-AdS}), by slowly 
increasing the parameter $\alpha$.
As a general remark, the qualitative features of all solutions we have constructed
are rather similar to the MPdS$_5$ case.
For $\alpha>0$, we have noticed only quantitative difference in the values on the cosmological horizon and
at infinity, for a number of parameters of interest.

 In order to limit the amount of numerical investigation, we have studied in  details  mainly the case
 $r_h=1$, $r_c=3$. For the non-rotating limit, this corresponds to $\ell^2=10$ and
 %\beq
 $ -(8  \pi G) E/V_3 =  {3}M/{4} =  {3(5\alpha+9)}/{40},~~4\pi T_H^h =  {8}/({5(\alpha+1)}).$
 %\eeq
  
The profiles of the metric functions of a typical EGB-dS rotating black hole solution
corresponding to $\alpha = 1$, $\Omega_H^h=0.66$, $\ell\simeq 3.2$
are presented on Figure 1. 
One can see that the rotation leads to non constant values for $h(r)/r^2$ and  $b(r)\neq f(r)$,
especially in the region close to the black hole horizon.
Also, the metric functions and their derivatives are continuous at the cosmological horizon (although
to simplify
the plot we presented there only the profile of $b'(r)$).

Several parameters characterizing the solutions  are represented in Figure 2  as a function of
 the angular velocity at the black hole horizon. 
 The data corresponding to MPdS$_5$ solution is represented by the dotted lines and
 results from analytical calculations. In contrast,
 the curves correponding to the EGB theory are represented by continuous lines and result from our numerical calculation.
(The energy $E$ and angular momentum $J$ are represented in the units of $V_3/(4\pi)$.
The Hawking temperature is represented in units $1/(4 \pi)$ and horizon area $A_H$ in units $V_3$, while we  have set also $G=1$ in all data.)
Along with the case of the MPdS$_5$ solutions, the EGB black holes
 exist up to a maximal value of $\Omega_H^h= \Omega_{max}$. 
 For $\alpha=0$ one finds
\be
    \Omega_{H(max)}^h= \sqrt{\frac{2}{r_c^2+2 r_h^2}} \ 
    \frac{r_c r_h^3(r_c^2+r_h^2)}{r_c^2 r_h^4+r_c^2+ 2r_h^2}  \ \ , \ \ 
      \frac{1}{\ell^2}= \frac{1}{r_c^2+2r_h^2}.
\ee

 %%%%%%%%%%%%%%%%%%%%%%%%%%%%%%%%%%%%%%%%%%%%%%%%%%%%%%%%%%%%%%%%%%%%%%%%%%%%%%%
\begin{figure}[ht]
\hbox to\linewidth{\hss%
	\resizebox{11cm}{8cm}{\includegraphics{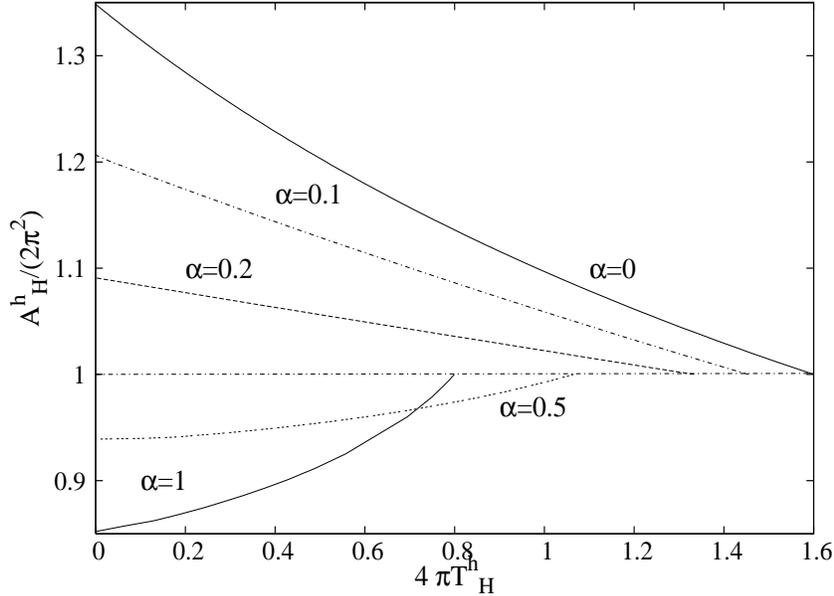}}
\hss}
 	\caption{The black hole event horizon area is plotted as function of the black hole temperature
 	for several values of the GB coupling constant $\alpha$.
 	These solutions have $r_h=1$, $r_c=3$ and $\ell^2=10$. }
\label{Fig2}
\end{figure}
%%%%%%%%%%%%%%%%%%%%%%%%%%%%%%%%%%%%%%%%%%%%%%%%%%%%%%%%%%%%%%%%%%%%%%%%%%%%%%%
For the cases we have investigated, when fixing the values of $r_h,r_c$,
this maximal value gets larger when the Gauss-Bonnet coupling constant 
 increases. 
 In this limit, 
the solution approaches an extremal black hole, $i.e.$  the functions
$f(r)$,$b(r)$ reach a double root at the black hole event horizon.  
However, these function still present a simple zero at the cosmological horizon.
For the values of the parameters adopted here we find $ e.g.$
$\Omega_H^h \approx 0.64$ for $\alpha = 0$ and $\Omega_H^h \approx 0.72$ for $\alpha = 1$.

 For the Einstein gravity black holes, the event horizon area $A_H^h$ increases with increasing $\Omega_H^h$
 while the Hawking temperature $T_H^h$ decreases. The entropy $S(T_H)$ turns out to be a 
 decreasing function of  the temperature, for fixed event horizon radius.  
 We have found that when the GB parameter $\alpha$ is large enough, the scenario is completely different. For instance, both 
 $T_H^h$ and $A_H^h$ decreases while $\Omega_H^h$ increases. The entropy is an increasing function of $T_H^h$.
 In Figure 3 we show this behaviour for several values of $\alpha$. There the parameter varying on the $A_H^h(T_H^h)$
curves is  $\Omega_H^h$. 

As with other rotating black holes, these solutions present also an ergoregion
inside of which the observers 
cannot remain stationary, and will move in the direction of rotation.
The ergoregion is the region bounded by the black hole event horizon, located 
at $r=r_h$ and the stationary limit surface, 
or the ergosurface, with $r=r_E<r_c$.
 The Killing 
vector $\partial/\partial t$ becomes null on the  ergosurface, $i.e.$ 
$g_{tt}= -b(r_E)+r_E^2 w(r_E)^2=0$. 
For the ansatz (\ref{metric}), the  ergosurface does not interesect the horizon. %
 We observe indeed that, for rotating solution with fixed $\Omega_H^h$, the value  $r_E$ decreases slightly
and get closer to $r_h$ when $\alpha$ increases. 
For exemple, with $\Omega_H^h = 0.66$, we get $r_E/r_h\approx1.34$ and $r_E/r_h \approx 1.27$
respectively for $\alpha = 0.1$ and $\alpha = 1$. 
In principle, there is also a second value of $r$,  
located outside the cosmological horizon, where the Killing 
vector $\partial/\partial t$ becomes null again.
 However, for all solutions we could construct,
 the metric component $g_{tt}$ there is dominated by $b(r)$
and thus the "cosmological" ergo-radius hardly differs from $r=r_c$.

Another qualitative difference between Einstein and EGB black holes resides in the
magnitude of the ratio $\rho \equiv f'/b'|_{r=r_h}$. For instance, for $\alpha=0$, we have $\rho<1$
for all values of the angular momentum. 
However, when $\alpha$ got sufficiently large,  one can find solutions with  $\rho>1$.

%%%%%%%%%%%%%%%%%%%%%%%%%%%%%%%%%%%%%%%%%%%%%%%%%%%%%%%%%%%%%%%%%%%%%%%%%%%%%%%%%%%%%%%%%%%%%%%%%%%%%%
\section{Further remarks}
%%%%%%%%%%%%%%%%%%%%%%%%%%%%%%%%%%%%%%%%%%%%%%%%%%%%%%%%%%%%%%%%%%
In  this work we have presented the boundary counterterm  
that removes the divergences of the action and conserved quantities of the solutions in 
EGB theory with a positive cosmological constant for a spacetime dimension $d\leq 7$.
Similar to the case of Einstein gravity, the counterterm is built up with
curvature invariants of  the boundary metric. 
Their coefficients, however, present an explicit dependence of the dimensionless factor $\alpha^2\Lambda$.

Here one should say that the expression of the counterterm  proposed in this paper
was obtained by demanding cancellation of divergencies for a number of solutions in EGB theory, which
was also the approach used
in initial work on the boundary counterterm in Einstein gravity
 \cite{Balasubramanian:1999re}, \cite{Emparan:1999pm}.
However, for asymptotically dS solutions in the Einstein gravity, there exist an algorithmic procedure
for constructing $I_{ct}$ in a rigurous way, and so 
its determination is unique for $\alpha=0$ \cite{Ghezelbash:2001vs}.
This procedure involves solving the Einstein equations (written in Gauss-Codacci form) in terms
of the extrinsec curvature functional of the boundary and its derivatives to isolate the divergent parts.
All divergent contributions are independent of 
the boundary normal and can be expressed in terms of intrinsic boundary data.
In principle, this approach can be extended to asymptotically  dS solutions 
 in EGB theory, the only obstacle we can see at this stage being the huge complexity of 
the required computation.
A more promising direction would be to look for the expression of $I_{ct}$
in the linear order in $\alpha$, by generalising the work in \cite{Liu:2008zf}
to the $\Lambda>0$  case.

For asymptotically AdS solutions, an alternative regularization prescription for 
any Lovelock theory has been proposed in \cite{kofinas}. 
This approch uses boundary terms with
explicit dependence on the extrinsic curvature $K_{ab}$,
also known as Kounterterms.
It would be interesting to generalize the approach in \cite{kofinas}
to dS asymptotics and to compare the results with those found here.

In the second part of this work, the general formalism
has been applied for several different
asymptotically  dS black hole solutions in EGB theory.
Apart from several known solutions, we have 
considered also rotating black holes with two equal-magnitude angular momenta
 in $d=4+1$ EGB theory with a positive cosmological
 constant.
 Although the numerical difficulties associated with the existence of a
cosmological horizon prevented us from a systematic study
of the parameter space, we have presented arguments for the existence
of nontrivial generalization in EGB theory of a  
 particular class of the known MPdS$_5$ black holes.

As avenue for future research, it would be interesting to consider the status 
of "{\it the maximal mass conjecture}" in
EGB theory, by using the mass definition proposed in this work.   
Formulated in \cite{Balasubramanian:2001nb}  for Einstein gravity, this conjecture states that any asymptotically
dS spacetime cannot have a mass larger than the pure dS case without inducing a cosmological singularity.
Here we mention only the fact that all rotating  black holes we have constructed  in Section 4
satisfy this conjecture.

The conserved charges of the rotating solutions in this paper have been evaluated
on a Euclidean surface at future timelike infinity.
In principle, by using the the results in Section 2,
a similar computation can be performed for a spatially finite boundary inside the
cosmological event horizon.
The corresponding problem for Kerr-dS rotating black holes in Einstein gravity
has been considered in ref. \cite{Dehghani:2002np}.
The results in that work show that quasilocal angular momentum
is independent on the radius of the boundary, which does not hold for
the total mass of the solutions.

The relevance  of the results discussed in this paper in a dS/CFT context is another
interesting open question.
For the  AdS/CFT case, the higher derivatives curvature terms can be viewed as the corrections of
large $N$  expansion of the  boundary CFT in the strong coupling limit, see $e.g.$ \cite{Fayyazuddin:1998fb}. 
For the asymptotically dS case, any progress in this direction is likely to require first a 
better understanding of the conjectured dS/CFT correspondence \cite{Hull:1998vg} with $\alpha=0$.
\\
\\
%%%%%%%%%%%%%%%%%%%%%%%%%%%%%%%%%%%%%%%%%%%%%%%%%%%%%%%%%%%%%%%%%%%%%%%%%
{\bf\large Acknowledgements} 
\\
Y. B. thanks  the Belgian FNRS for financial support.  
The work of ER was supported by a fellowship from the Alexander von Humboldt Foundation. 
%%%%%%%%%%%%%%%%%%%%%%%%%%%%%%%%%%%%%%%%%%%%%%%%%%%%%%%%%%%%%%%%%%%%%%%%%%

%%%%%%%%%%%%%%%%%%%%%%%%%%%%%%%%%%%%%%%%%%%%%%%%%%%%%%%%%%%%%%%%%%%%%%%%%%%%%%%%%%%%%  
 \begin{small}
 
%%%%%%%%%%%%%%%%%%%%%%%%%%%%%%%%%%%%%%%%%%%%%%%%%%%%%%%%%%%%%%%%%%%%%%%%%%%%%%
 \end{small}

\end{document}